%% file: talk2.tex
\documentstyle[sprocl,epsfig]{article}

\newcommand{\beq}{\begin{equation}}
\newcommand{\eeq}{\end{equation}}
\newcommand{\bea}{\begin{eqnarray}}
\newcommand{\eea}{\end{eqnarray}}
\newcommand{\barr}{\begin{array}}
\newcommand{\earr}{\end{array}}
\newcommand{\bc}{\begin{center}}
\newcommand{\ec}{\end{center}}
\newcommand{\btab}{\begin{tabular}}
\newcommand{\etab}{\end{tabular}}

\newcommand{\nn}{\nonumber}
\newcommand{\ra}{\rightarrow}

\newcommand{\dro}{\Delta\rho}

\newcommand{\al}{\alpha}

\newcommand{\G}{\Gamma}
\newcommand{\Gmu}{G_{\mu}}

\newcommand{\ganu}{\gamma_{\nu}}

\newcommand{\gafi}{\gamma_5}

\newcommand{\noi}{\noindent}
\newcommand{\epmf}{e^+e^- \rightarrow f\bar{f}}

\newcommand{\epm}{e^+e^-}

\newcommand{\sm}{Standard Model }

\newcommand{\dal}{\Delta\alpha}

\newcommand{\mz}{M_Z^2}
\newcommand{\mw}{M_W^2}

\newcommand{\Dr}{\Delta r}

\newcommand{\alr}{A_{\rm LR}}
\newcommand{\afb}{A_{\rm FB}}

\newcommand{\ass}{asymmetries }

\newcommand{\prd}{{\it Phys.\ Rev.\ }}
\newcommand{\zp}{{\it Z.\ Phys.\  }}
\newcommand{\plb}{{\it Phys.\ Lett.\ }}

\newcommand{\prl}{{\it Phys.\ Rev.\ Lett.\ }}
\newcommand{\np}{{\it Nucl.\ Phys.\ }}

\newcommand{\tht}{\theta_{\tilde{t}}}
\newcommand{\sbottom}{\tilde{b}}
\newcommand{\sbl}{\tilde{b}_L}
\newcommand{\sta}{\tilde{t}_1}
\newcommand{\stb}{\tilde{t}_2}
\newcommand{\rbsg}{R_{b\rightarrow s \gamma}}
\begin{document}

\title{ELECTROWEAK PRECISION OBSERVABLES IN THE MSSM AND GLOBAL
       ANALYSES OF PRECISION DATA}

\input titlepage

\author{ W. HOLLIK }

\address{ Institut f\"ur Theoretische Physik, Universit\"at Karlsruhe \\  
          D-76128 Karlsruhe, Germany \\
          E-mail: Wolfgang.Hollik@physik.uni-karlsruhe.de}


\maketitle\abstracts{
The calculation of electroweak precision observables in the MSSM
is reviewed. The description of the 1997 data
and the results of updated global fits are discussed
in comparison with the Standard Model.}

\section{Introduction}
Experiments have measured the electroweak observables 
of the $W$ and $Z$ bosons \cite{lep}
and the top quark \cite{top} with an impressive
accuracy thus providing precision tests of the electroweak theory
at a level which had never been reached before.
The description of the current precision data by the minimal
Standard Model is extaordinarily successful, with a few observed
deviations of the order  $\geq 2  \sigma$. They lower the
quality of the \sm fits, but they can be understood as
fluctuations which are statistically normal.

\smallskip
Extensions of the \sm hence are essentially theoretically motivated.
Among the supersymmetric extensions, the $R$-parity conserving 
minimal supersymmetric standard
model (MSSM) with the minimal particle content 
plays a special role 
as the most predictive framework beyond the Standard Model.
Besides introducing 
the superpartners for the standard particles, the Higgs sector 
has to be augmented by a second scalar doublet. 
The structure of the MSSM allows a similarly complete calculation of
the electroweak precision observables
as in the Standard Model in terms of one Higgs mass
 and the ratio $v_2/v_1$ of the Higgs vacua
together with the set of
SUSY soft breaking parameters fixing the chargino/neutralino and
scalar fermion sectors.
Besides the direct searches for SUSY particles, which as yet
have not been successful \cite{limits}, the interest for
indirect tests of the MSSM
in terms of virtual effects from the non-standard particles  
has triggered a broad activity 
in calculating and investigating the supersymmetric quantum effects
in the electroweak precision observables  
\cite{higgs,susy1,susydelr,susy3,deboer}. 
Complete one-loop calculations 
are available for the quantity
$\Delta r$ in the $M_W$-$M_Z$ correlation \cite{susydelr} 
and for the $Z$ boson observables
\cite{susy3,deboer} and have recently been improved by the 
supersymmetric QCD corrections to the $\rho$-parameter
at the two-loop level \cite{drosusy}.

\section{MSSM entries}
The quantum contributions to the electroweak precision observables 
contain, besides the standard input, the following
entries:

\medskip \noi
{\it Higgs sector:}
The scalar sector of the MSSM, consisting of three neutral particles
$h^0,\, H^0, \, A^0$ and a pair of charged bosons $H^{\pm}$, 
is completely determined by the value
of $\tan\beta=v_2/v_1$ and the pseudoscalar mass $M_A$, together with the 
radiative corrections. The latter ones can be easily 
taken into account by means of 
the effective potential approximation with the leading terms $\sim m_t^4$,
including mixing in the scalar top system \cite{ellisetal}.
In this way, the coupling constants of the various Higgs particles
to gauge bosons and fermions can be taken from \cite{hunter}
substituting only the scalar mixing angle $\alpha$ by the improved 
effective mixing angle which is obtained from the diagonalization of the
scalar mass matrix. The mass of the light $h^0$ is constrained
to $< \, 130$ GeV when the dominant two-loop corrections are included
\cite{hoang}.

\medskip \noi
{\it Chargino/Neutralino sector:}
The chargino (neutralino) masses and the mixing angles in the gaugino
couplings are calculated from soft breaking parameters $M_1$, $M_2$ and
 $\mu$ in
the chargino (neutralino) mass matrix \cite{hunter}. 
For practical calculations, 
the GUT relation $M_1 = 5/3 \tan^2 \theta_W \, M_2$ is
conventionally  assumed. 

\smallskip \noi
The chargino $2 \times 2$ mass matrix is given by
\begin{equation}\label{charmat}
  \cal{M}_{\rm \tilde{\chi}^\pm} \rm = \left( \begin{array}{l l}
    M_2 & \quad M_W \sqrt{2} \sin \beta \\ M_W \sqrt{2} \cos \beta &
      \quad  - \mu \\
    \end{array}  \right) \ ,
\end{equation} \\
with the SUSY soft breaking parameters $\mu$ and $M_2$ in the diagonal
matrix elements. The physical chargino mass states $\tilde{\chi}^{\pm}_i$ 
are the rotated
wino and charged Higgsino states:
\begin{eqnarray}
\tilde{\chi}^+_i & = & V_{ij} \psi^+_j   \nonumber \\
\tilde{\chi}^-_i & = & U_{ij} \psi^-_j  \ ; \ i,j = 1,2  \ .
\end{eqnarray}
$V_{ij}$ and $U_{ij}$ are unitary chargino mixing matrices obtained from
the diagonalization of the mass matrix Eq.~(\ref{charmat}):
\begin{equation}
\rm U^* \cal{M}_{\rm \tilde{\chi}^\pm} \rm  V^{-1} = diag(m_{\tilde{\chi}^\pm_1},m_{\tilde{\chi}^\pm_2}) \ .
\end{equation}

\smallskip \noi
The neutralino $4 \times 4 $ mass matrix 
$\cal{M}_{\rm \tilde{\chi}^0}$ 
can be written as:   
\begin{equation}\label{neutmat}
 \left( \begin{array}{cccc}
 M_1 & 0 & - M_Z \sin \theta_W \cos \beta & M_Z \sin \theta_W \sin \beta \\
 0 & M_2 & M_Z \cos \theta_W \cos \beta & - M_Z \cos \theta_W \sin \beta \\
- M_Z \sin \theta_W \cos \beta & M_Z \cos \theta_W \cos \beta & 0 &  \mu \\
 M_Z \sin \theta_W \sin \beta & - M_Z \cos \theta_W \sin \beta & \mu & 0
 \\  \end{array} \right) 
\end{equation}
where the diagonalization can be obtained by the unitary matrix $N_{ij}$: 
\begin{equation}
  \rm N^* \cal{M}_{\rm \tilde{\chi}^0} \rm N^{-1}  = diag(
  m_{\tilde{\chi}_i^0}) \ .
\end{equation}

\smallskip \noindent
The elements $U_{ij}$, $V_{ij}$, $N_{ij}$ of the diagonalization
matrices  enter the couplings of the 
charginos, neutralinos and sfermions to fermions and gauge bosons, as
explicitly given in ref.~\cite{hunter}. Note that the sign convention on the
parameter $\mu$ is opposite to that of ref.~\cite{hunter}.

\bigskip \noi 
{\it Sfermion sector:}
The physical masses of squarks and sleptons are given by the eigenvalues
of the $2 \times 2$ mass matrix: \hspace{1cm}
$ \cal M_{\rm \tilde{f}}^{\rm 2} \quad  = $
\begin{equation}
  \left( \begin{array}{ll}
 M_{\tilde{Q}_L}^2 + m_f^2 + M_Z^2 (I_3^f - Q_f s_W^2) \cos 2 \beta &
 \quad  m_f (A_f + \mu \{ \cot \beta , \tan \beta \} ) \\
 m_f (A_f + \mu \{ \cot \beta , \tan \beta \} ) &  \quad
 M_{\{\tilde{U},\tilde{D}\}_R}^2 + m_f^2 + M_Z^2 Q_f s_W^2 \cos 2 \beta
 \end{array} \right) \ ,
\label{sqmatrix}
\end{equation}
with SUSY soft breaking parameters $M_{\tilde{Q}_L}$,
$M_{\tilde{U}_R}$, $M_{\tilde{D}_R}$, $A_f$, and $\mu$.
It is convenient to use  the following notation for the 
off-diagonal entries in Eq.~(\ref{sqmatrix}):
\begin{equation}
  M_f^{LR}  = A_f + \mu \{ \cot \beta , \tan \beta \} \ .
\label{glaprime}
\end{equation}
Scalar neutrinos appear only as left-handed mass eigenstates.
Up and down type sfermions in (\ref{sqmatrix}) are distinguished by
setting f=u,d and the corresponding 
$\{u,d\}$ entries in the parenthesis.
Since the non-diagonal terms are proportional to $m_f$, it seems
natural to assume unmixed sfermions for the lepton and quark case
except for the scalar top sector.
The $\tilde{t}$ mass matrix is diagonalized by a rotation matrix with
a mixing angle $\tht$. 
Instead of $M_{\tilde{Q}_L}$, $M_{\tilde{t}_R}$, $M_{\tilde{b}_R}$, 
$M_t^{LR}$
for the $\tilde{b}$, $\tilde{t}$ system ($Q=t,b$)
the physical squark masses
$m_{\tilde{b}_L}, m_{\tilde{b}_R}$, $m_{\tilde{t}_2}$ can be used
together with 
$M_t^{LR}$ or, alternatively, the stop mixing angle $\tht$.
For simplicity one may 
assume $m_{\tilde{b}_L}=m_{\tilde{b}_R}$, and 
$\tilde{u}$, $\tilde{d}$, $\tilde{c}$, $\tilde{s}$
to have masses equal to the 
$\tilde{b}$ squark mass.
The labels for the $\tilde{t}$ mass eigenstates are chosen in such a way
that one gets $\sta= \tilde{t}_L, \; \stb = \tilde{t}_R$ for the
case of zero mixing.

\section{Precision observables}
\subsection{The $\rho$-parameter}
A possible mass splitting between the left-handed components of the 
 $\tilde{b}$ and $\tilde{t}$ system   
yields a contribution to the $\rho$-parameter
$\rho = 1+ \Delta\rho$
in terms of  
(neglecting left-right mixing in the $\sbottom$ sector)
\bea   
\label{deltarho}
\Delta \rho_{\tilde{t}\tilde{b}} & = & \frac{3 \Gmu}{8 \pi^2 \sqrt{2}} \,
\left[ \cos^2\tht\; F_0(m_{\sta}^2,m_{\sbl}^2) 
        + \sin^2\tht \;  F_0(m_{\stb}^2,m_{\sbl}^2)    \right.  \nn \\
    & &  \left.
 - \sin^2\tht \cos^2\tht \; F_0(m_{\sta}^2,m_{\stb}^2) \right] \nn \\
 {\rm with} & & F_0(x,y) = x+y-\frac{2xy}{x-y} \log \frac{x}{y} 
\eea
%
%

\begin{figure}[htb]
\vspace{-5cm}
\centerline{
\epsfig{figure=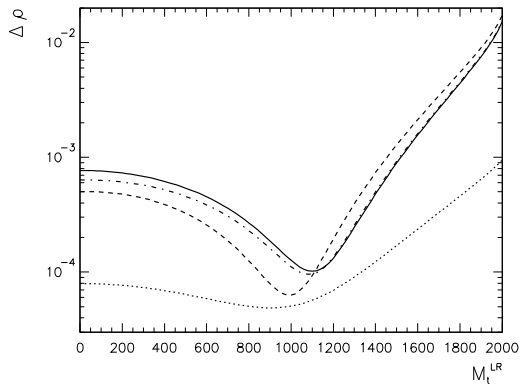,height=15cm}}
\vspace{-5cm}
\caption{$\Delta\rho_{\tilde{t}\tilde{b}}$ for $\tan\beta=1.6$.
         $M_{\tilde{t}_L}/M_{\tilde{t}_R} = 1000/300$ (dotted),
           300/1000 (solid)}
\label{rho1}
\end{figure}

\smallskip \noi
Examples for $\Delta\rho$ from the squark loops of the third generation,
which can become remarkably large,
are displayed in Figure \ref{rho1}.
 Recently the 2-loop $\al_s$ corrections have been
computed \cite{drosusy,heinemeyer}
which can amount to a sizeable fraction  of the 1-loop
$\Delta \rho_{\tilde{b}\tilde{t}}$.
As a universal loop contribution, $\dro_{\tilde{t}\tilde{b}}$
enters the quantity $\Dr$ and the
$Z$ boson couplings through the relations 
(\ref{deltar}) and (\ref{rhoef})
and is thus significantly constrained by the data on $M_W$ and the
leptonic $Z$ widths.

\subsection{The vector boson masses}
The correlation between
the masses $M_W,M_Z$ of the vector bosons          in terms
of the Fermi constant $\Gmu$  is given by
\bea
\label{deltar}
\frac{\Gmu}{\sqrt{2}} &  = &
            \frac{\pi\al}{2s_W^2 M_W^2} \cdot \frac{1}
        {1- \Dr}      \nn \\
       & = &      \frac{\pi\al}{2s_W^2 M_W^2} \cdot \frac{1}
             {(1-\Delta\al)\cdot
          (1+\frac{c_W^2}{s_W^2}\dro) \, -\,(\Dr)_{\rm rem}}  \\[0.2cm]
{\rm with}\quad s_W^2 & = & 1- \frac{\mw}{\mz} \, .  \nn
\eea
Therein, $\dal = 0.0595\pm 0.0007$ \cite{jeger} is the QED vacuum
polarization of the photon from the light fermions, $\dro$ is the 
irreducible quantum contribution to the $\rho$-parameter, and
$(\Dr)_{\rm rem}$ contains the residual loop terms.
Complete 1-loop calculations for 
the quantity $\Dr$ have been performed in ref's \cite{susydelr}.
Besides the higher order contributions beyond 1-loop from the Standard
Model (see \cite{hollik} for more information and references), the
SUSY-QCD 2-loop terms to $\dro$ are meanwhile available, as mentioned above
in Section 3.1,  as well as the complete 2-loop gluonic QCD corrections to 
the squark loop contributions to $\Dr$ \cite{heinemeyer}. The SUSY 
QCD corrections to $\Dr$ are well approximated by the SUSY
QCD corrections to $\dro$. The 2-loop terms correspond to a shift in
the $W$ mass of 10-20 MeV  for not too heavy SUSY partners of the quarks
and are thus of phenomenological importance.

\begin{figure}[htb]
\vspace{-2cm}
\centerline{
\mbox{\epsfxsize8.0cm\epsffile{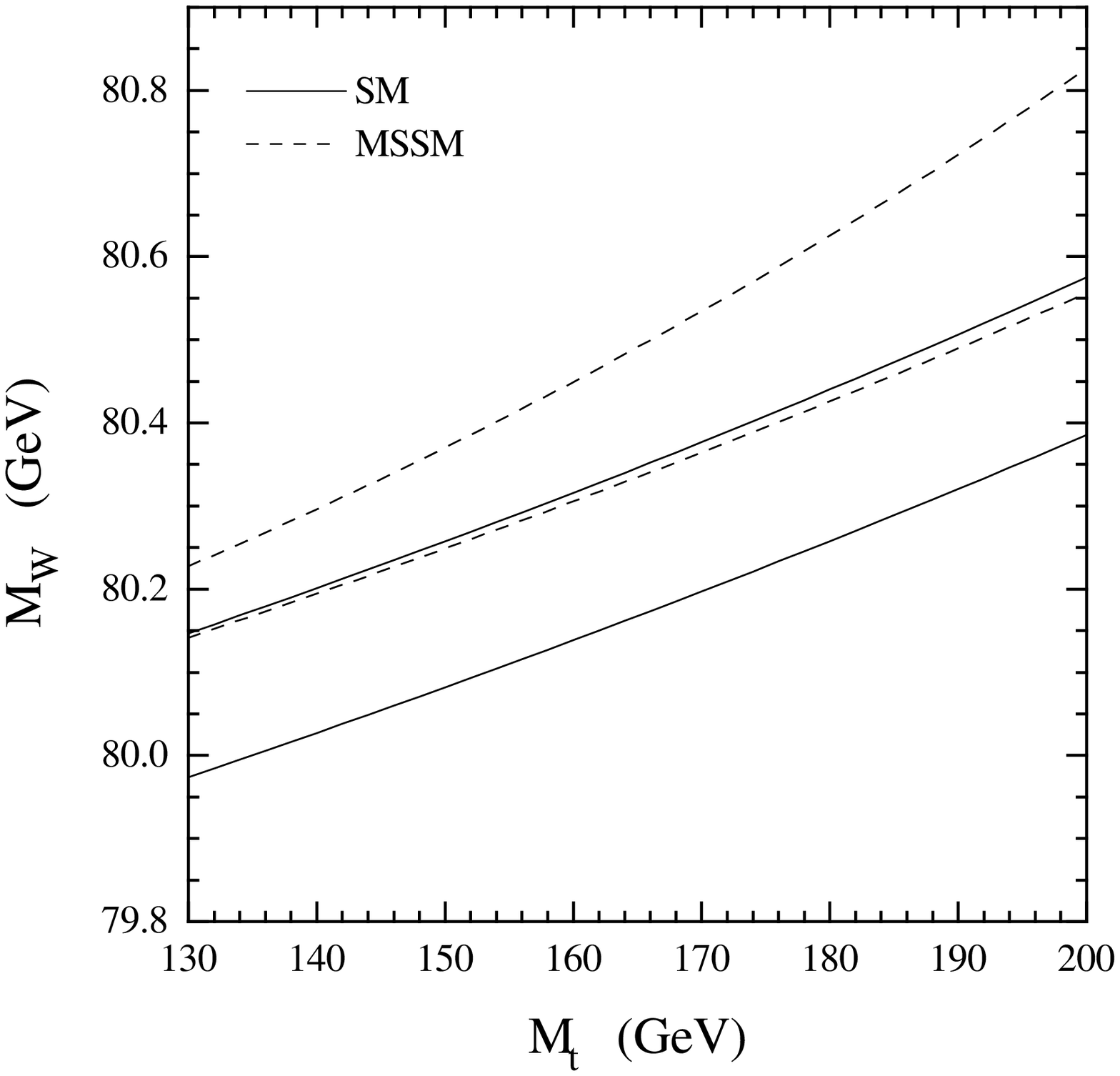}}} 
\vspace{-3cm}
\caption{The $W$ mass range in the Standard Model (-----) and the
         MSSM (- - -). Bounds are from the non-observation of Higgs
         bosons and SUSY particles at LEP2.} 
\label{susymw}
\end{figure}

\medskip
By Eq.~(\ref{deltar}) the value for $M_W$ is fixed  when $M_Z$
is taken as input and the MSSM model parameters have been chosen.
Figure \ref{susymw}
displays the range of predictions for $M_W$ in the Standard Model (SM)
and in the MSSM. Thereby it is assumed that no direct discovery has been
made at LEP2 for constraining the model parameters.
As one can see, precise determinations of $M_W$ and $m_t$
can become decisive for the separation between the  models.
The present world average \cite{lep}
 $M_W = 80.43\pm 0.08$ GeV 
together with \cite{top} $m_t=175.6\pm 5.5$ GeV
shows a slight, but not significant,
preference for the MSSM.

\setlength{\unitlength}{0.7mm}

\subsection{$Z$ boson observables}
With $M_Z$ as a precise input parameter, 
the predictions for the partial widths
as well as for the asymmetries
can conveniently be calculated in terms of effective neutral
current coupling constants for the various fermions entering
the neutral current vertex as follows:
\bea
\label{nccoup}
   J_{\nu}^{\rm NC}      &  =  &  
  g_V^f \,\ganu -  g_A^f \,\ganu\gafi     \nn \\ 
  & = &   \left( \rho_f \right)^{1/2}
\left[ (I_3^f-2 Q_f\, s_f^2)\ganu-I_3^f\ganu\gafi \right] 
\eea
with form factors 
$\rho_f$ for the overall normalization and  with  the
effective mixing angles $s_f^2 = \sin^2 \theta_f$ for the corresponding
fermions. 
The leading universal 
terms associated with the $\rho$-parameter enter the
$Z$ boson couplings through 
\bea
\label{rhoef}
 \rho_f & = &  \frac{1}{1-\dro} + \cdots    \nn  \\
  s_f^2 & = & s_W^2 + c_W^2\, \dro + \cdots  
\eea
Large non-universal contributions from the Higgs and the genuine
SUSY particle sector are possible especially for the $b$-quark
couplings.

\smallskip \noi
{\it Asymmetries and mixing angles:}

\noi 
The effective mixing angles are of particular interest since
they determine the on-resonance asymmetries via the combinations
   \beq
    A_f = \frac{2g_V^f g_A^f}{(g_V^f)^2+(g_A^f)^2}  \, .
\eeq
Measurements of the \ass hence are measurements of
the ratios
\beq
  g_V^f/g_A^f = 1 - 2 Q_f s_f^2
\eeq
or of the effective mixing angles $s^2_f$.

\smallskip \noindent
The measurable quantities are: 

\smallskip \noi
$-$ the forward-backward asymmetries in $\epmf$:
$$ \afb^f =\frac{3}{4} A_e \cdot A_f $$
$-$ the left-right asymmetry:
$$ \alr = A_e $$
$-$ the $\tau$ polarization in $\epm \ra \tau^+\tau^-$ :
$$ P_{\tau} = A_{\tau} \, .$$
It is interesting to note that the non-standard loop contributions
to the leptonic mixing angle $s_e^2$ diminish the Standard Model value.
A small experimental value, as observed in $\alr$, could 
therefore be accomodated in the MSSM
also for a relatively high top mass.

\smallskip
\paragraph{\it $Z$ widths and cross sections:}
 
The total
$Z$ width $\Gamma_Z$ can be calculated 
essentially as the sum over the fermionic partial decay widths:
\beq
 \Gamma_Z = \sum_f \, \Gamma_f, \;\;\;\;
 \Gamma_f = \Gamma  (Z\ra f\bar{f}) \, .
\eeq
The peak cross sections for $\epmf$ (had) are determined by
\beq
  \sigma_0^f = \frac{12 \pi}{\mz} \frac{\G_e \G_f}{\G_Z^2} , \quad
 \sigma_0^{\rm had} = \frac{12 \pi}{\mz} \frac{\G_e \G_{\rm had}}{\G_Z^2} \, .
\eeq
 The partial widths, 
expressed in terms of the effective coupling constants,
read up to 2nd order in the fermion masses:
\bea
\Gamma_f
  & = & \G_0
 \, \left(
     (g_V^f)^2  +
     (g_A^f)^2 (1-\frac{6m_f^2}{\mz} )
                           \right)
 \cdot   (1+ Q_f^2\, \frac{3\al}{4\pi} ) 
          + \Delta\G^f_{QCD} \nn
\eea
with
$$
\G_0 \, =\,
  N_C^f\,\frac{\sqrt{2}\Gmu M_Z^3}{12\pi},
 \;\;\;\; N_C^f = 1
 \mbox{ (leptons)}, \;\; = 3 \mbox{ (quarks)}.
$$
and the QCD corrections  $ \Delta\G^f_{QCD} $
 for quark final states, both from standard (see \cite{hollik} for
references) and SUSY QCD.
The gluino contribution, however, is very small.

\begin{figure}[htb]
\centerline{
\epsfig{figure=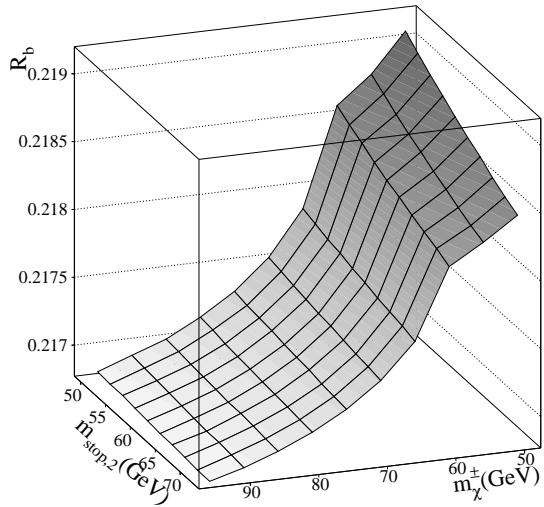,height=10cm}}
\vspace{-1.5cm}
\caption{$R_b$ in the light stop-chargino plane. $\tan\beta = 1.6$. 
 {\protect\cite{deboer}}   }
\label{rblow}
\end{figure}

\begin{figure}[htb]
\centerline{
\epsfig{figure=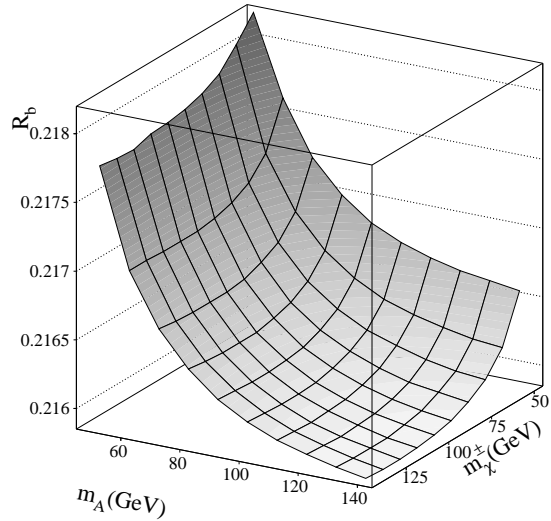,height=10cm}}
\vspace{-1.5cm}
\caption{$R_b$ in the pseudoscalar Higgs - chargino plane.
  $\tan\beta = 34$. {\protect\cite{deboer}} }
\label{rbhigh}
\end{figure}

\smallskip \noi 
Of particular interest is the quantity 
\beq
  R_{\rm b} = \frac{\G_{\rm b}}{\G_{\rm had}}
\eeq
which in the Standard Model is most sensitive to the mass of the top 
quark. 
As already known for  quite some time
\cite{higgs},
light non-standard
Higgs bosons for large $\tan\beta$ as well as light stop and charginos
influence the $b$-quark couplings remarkably and 
predict larger values for the ratio $R_b$ in the MSSM
 \cite{susy1,susy3}
(see figures \ref{rblow}, \ref{rbhigh}).

\clearpage

\section{MSSM global fits to precision data}
For obtaining the optimized SUSY parameter set a global
fit to all the electroweak precision data, including the top
mass measurement, has to be  performed.
Two strategies can be persued: (i) the parameters of the MSSM are
further restricted by specific model assumptions like unification
scenarios, or (ii) the MSSM parameters are considered as free 
quantitites chosen in the optimal way for improving the observed
deficiencies of the Standard Model in describing the data. 
(ii) has been applied in
\cite{deboer}, recently updated by Schwickerath \cite{schwick}.
In the past, the experimental value of $R_{\rm b}$ was $3.7 \sigma$
away from the Standard Model predictions and could be accomodated
much better in MSSM scenarios with light stop and charginos and/or
light scalar amd pseudoscalar Higgs bosons. In the meantime, $R_{\rm b}$
has decreased and is now much closer to the Standard Model value.
Instead, $\afb^{\rm b}$ shows a $2 \sigma$ deviation between data
and Standard Model prediction \cite{lep}.

\newpage

\begin{figure}[htb]
\centerline{
\epsfig{figure=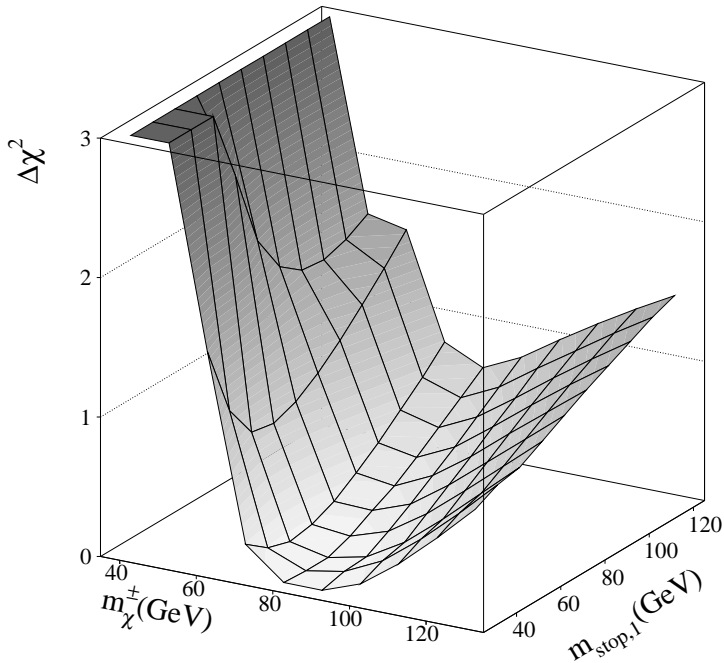,height=10cm}}
\caption{$\Delta\chi^2$ for $\tan\beta  = 1.5$. No limits for 
supersymmetric particle masses are imposed (from
 {\protect\cite{deboer}} updated in
 {\protect\cite{schwick}} )   }
\label{chi2}
\end{figure}

In order to obtain the best MSSM fits
the electroweak precision data
are taken into account together with the
error correlations \cite{lep}, and
the measurement of the branching ratio
$\rbsg = {\rm BR}(B\ra X_s \gamma)$ 
by CLEO and ALEPH \cite{cleo} with the
combined value $ (2.55\pm 0.61)\cdot 10^{-4}$. The $b \ra s\gamma$ rate
becomes too small if $\tan\beta$ is near one. For $\tan\beta =1.6$ and 35,
MSSM solutions can be found which are compatible with $\rbsg$ and the
electroweak precision data.
For the MSSM calculation of $\rbsg$, 
the results of \cite{barbieri} are used 
with a chosen renormalization scale of $\mu=0.65 m_b$.

\bigskip
For reducing the large number of parameters, the simplifying
assumptions described in Section 2 are applied. As a common 
squark mass scale the value $M_{\tilde{Q}} = 1$ TeV is chosen; 
$M_{\tilde{t}_R}$ and
the stop mixing angle $\tht$ are kept free to allow for a light stop state
$\stb$. This state  has to be 
predominatly right-handed to avoid deteriorations of the
$\rho$-parameter according to Eq.~(\ref{deltarho}).
In the slepton sector, common masses of 0.5 TeV are assumed  for both
left and right-handed components neglecting left-right mixing. 
The fits are insensitive to the detailed values of the heavy particles.
Figure \ref{chi2} shows the increase of the 
$\chi^2$ of the fit \cite{schwick} in the 
light stop-chargino plane ($\Delta\chi^2 = \chi^2-\chi^2_{\rm min}$). 
In each point, the parameters $m_t, \al_s, \tht, M_2$ are optimized.

\begin{table}[htb]
\caption{Fitted parameters and mass spectrum for the best 
         low and high $\tan\beta$ fits
         {\protect\cite{schwick}}, updated from
         {\protect\cite{deboer}}.
         Parameter values in
         brackets are not fitted, 
         but fixed to the given value during the fit.}
\vspace{0.4cm}
  \begin{center}
    \begin{tabular}{|l|c|c|}
      \hline
      &$\tan\beta$=1.5 & $\tan\beta$=35 \\
      \hline
$\chi^2$/d.o.f. &15.0/13 & 15.8/13\\
prob.           &   31\% &  26\%\\
      \hline
      \multicolumn{3}{|c|}{fitted parameters}\\
      \hline
$\alpha_s(M_Z^2)$         &0.1161 & 0.1196\\
$m_t$ [GeV]               &174.0  & 174.3\\ 
$M_2$ [GeV]               &68     & (1500) \\
$\mu$ [GeV]               &79     & 92 \\
$m_A$ [GeV]               & (1500)     & 60\\
$m_{\tilde{t}_2}$ [GeV]   & 80    & 81\\
$\tht$                    &-0.139 & 0.0126\\
\hline
\multicolumn{3}{|c|}{mass spectrum}\\
\hline
$\tilde{\chi_1}^\pm$,$\tilde{\chi_2}^\pm$ [GeV] &89, 126 & 92, 1504\\ 
$\tilde{\chi_1}^0$,$\tilde{\chi_2}^0$[GeV] &38,70 & 89,96\\ 
$\tilde{\chi_3}^0$,$\tilde{\chi_4}^0$[GeV] &115,123 & 716,1504\\ 
$m_{\tilde{t}_1}$ [GeV]  & 1023 & 1012\\
$M_h,M_H $ [GeV] & 98,1503 & 60,103 \\
$M_{H^\pm}$ [GeV] & 1502 & 124\\
$M_W$ [GeV]       & 80.42     &  80.43  \\
\hline
    \end{tabular}
  \end{center}
\label{results}
\vspace{0.5cm}
\end{table}

\medskip
Constraints from direct searches for SUSY particles \cite{limits}
can be taken into 
account with the help of a penalty function (see \cite{schwick}):
$M_A> 50$ GeV, $M_h > 60$ GeV, $m_{\tilde{\chi}^+} > 85$ GeV, 
$m_{\stb} > 80$ GeV. With these constraints the best fit results 
for a low and high $\tan\beta$ scenario are listed in Table
\ref{results}.  
The values of the strong coupling constant $\al_s$
is close to the best fit value of the Standard Model \cite{lep}
$\al_s ({\rm SM}) = 0.120 \pm 0.003$. 
For the low $\tan\beta$ scenario $\al_s$ is slightly lower; this is due
to the positive loop contribution to the partial
$Z$ width $\G_{\rm b}$, which is less pronounced in the high $\tan\beta$
scenario because of the constraints to the Higgs masses.
The particle spectrum for the best fits
suggests that some SUSY particles could be within  reach of LEP 2;
the $\chi^2$
in the region of the best low $\tan\beta$ fit, however,  increases
only slowly for increasing chargino masses.
Within the high $\tan\beta$ scenario both  neutral $h^0,A^0$ 
bosons are light and have practically the same mass.
This scenario can be excluded if no Higgs bosons
will be found at LEP 2.

\setlength{\unitlength}{0.7mm}
\begin{figure}[hbt]
\centerline{
\mbox{\epsfxsize8.0cm\epsffile{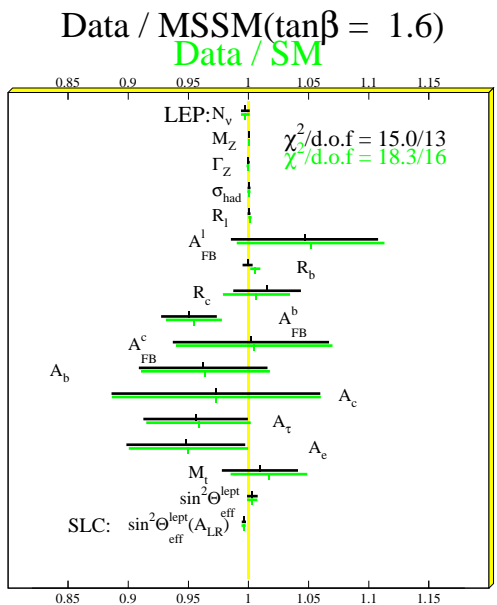}}} 
\vspace{-1cm}
\caption{Experimental data normalized to the best fit results in
         the SM and MSSM for low $\tan\beta$, 
         with experimental error bars 
         (updated from {\protect\cite{deboer}} in
         {\protect\cite{schwick}}).
         }
\label{mssm1}
\end{figure}

\setlength{\unitlength}{0.7mm}
\begin{figure}[hbt]
\centerline{
\mbox{\epsfxsize8.0cm\epsffile{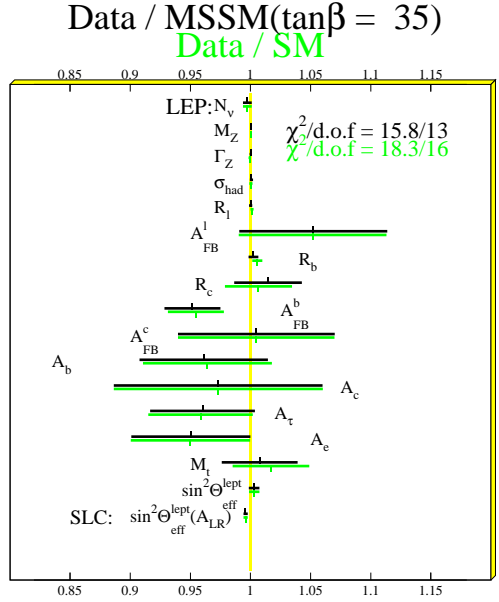}}} 
\vspace{-1cm}
\caption{Experimental data normalized to the best fit results in
         the SM and MSSM for high $\tan\beta$, 
         with experimental error bars 
         (updated from {\protect\cite{deboer}} in
         {\protect\cite{schwick}}).
         }
\label{mssm2}
\end{figure}

A direct comparison to the Standard Model fit is shown  in
Figure \ref{mssm1} for $\tan\beta=1.6$. A very similar plot is obtained for
$\tan\beta=35$ displayed in Figure \ref{mssm2}.
The present small difference between the experimental and the SM value of 
$R_{\rm b}$ can be completely removed in the MSSM for the low $\tan\beta$.
Other quantities
are practically unchanged; in particular, the difference between data
and theory observed in $\afb^{\rm b}$ cannot be significantly diminished.
The $\chi^2$ of the MSSM fits are lower than in the
Standard Model, but due to the larger number of parameters the
probability in the MSSM
is not  higher than in the Standard Model (31\%); for high $\tan\beta$ the
probability is even lower.
(The small difference to the SM result of the LEP Electroweak Working Group
is due to the inclusion of $\rbsg$ in our fit.)
This means that the MSSM yields an equally good description of the data
as the minimal Standard Model 
and is thus competitive, but it does not appear as 
 superior.

\section{Conclusions}
The supersymmetric extension of the \sm realized in terms of the MSSM
permits a similar complete calculation of the electroweak precision
observables as in the Standard Model. The version with all SUSY
particles and the $A^0, H^0, H^{\pm}$ Higgs bosons heavy  is
equivalent to the \sm with a light Higgs around 120 GeV. 
The possibility of having some of the non-standard particles light,
in the mass range around 100 GeV, opens the option to remove some of
the observed deviations between data and theory, like in $R_{\rm b}$.
The overall $\chi^2$ of a global fit is therefore slightly lower than
in a corresponding \sm fit; the larger number of parameters,
however, reduce the $\chi^2$ per degree of freedom and thus the
probability of the MSSM to that of the Standard Model. 
The MSSM thus yields an equally good description of the precision data
as the Standard Model.
It is always possible to worsen the fits by inappropriate choices of
the model parameters, which can be used to exclude certain regions
of the parameter space, as done recently in \cite{erler}.
The $W$ mass predicted by the MSSM is always higher
than in the Standard Model; present data show a slightly better agreement
with the MSSM. 
From the increasing experimental accuracy
for $M_W$ and $m_t$ 
in the future one hence expects a very  important probe
for both the \sm and the MSSM.

\section*{Acknowledgements}
I want to thank Joan Sol\`a and the Organizing Committee for the 
invitation to this Workshop and 
for the kind hospitality at the
Universitat Autonoma de Barcelona.
Special thanks are extended to Ulrich Schwickerath
who provided me with his results 
on the updated MSSM fits.

\end{document}

%% file: titlepage.tex
\thispagestyle{empty}
\setcounter{page}{0}
\def\thefootnote{\fnsymbol{footnote}}

\begin{flushright}
KA--TP--23--1997\\
\end{flushright}

\vspace{1cm}

\begin{center}

{\large\sc {\bf Electroweak Precision Observables in the MSSM}}

\vspace*{0.4cm} 

{\large\sc {\bf and Global Analyses of Precision Data}}
\footnote{talk given at the International Workshop on
          Quantum Effects in the MSSM, Barcelona,
          9 - 13  September 1997}

\vspace{1cm}

{\sc W. Hollik
}

\vspace*{1cm}

     Institut f\"ur Theoretische Physik \\ Universit\"at Karlsruhe \\
     D-76128 Karlsruhe, Germany

\vspace*{0.4cm}


\end{center}

\vspace*{1cm}

\begin{abstract}
The calculation of electroweak precision observables in the MSSM
is reviewed. The description of the data of 1997 
and  results of updated global fits are discussed
in comparison with the Standard Model.
\end{abstract}

\def\thefootnote{\arabic{footnote}}
\setcounter{footnote}{0}

\newpage

%% file: talk2.bbl
\section*{References}
\begin{thebibliography}{99}
%
\bibitem{lep}
The LEP Collaborations ALEPH, DELPHI, L3, OPAL, the LEP Electroweak
Working Group and the SLD Heavy Flavor Working Group,
CERN-PPE/96-183 (1996); for the update see: \\
          J. Timmermans, Proceedings of the
           {\it XVIII International Symposium on 
           Lepton-Photon Interactions}, Hamburg 1997; \\ 
           D. Ward,  Proceedings of the {\it International Europhysics
           Conference on High Energy Physics}, Jerusalem 1997
\bibitem{top} P. Giromini,  Proceedings of the
           {\it XVIII International Symposium on 
           Lepton-Photon Interactions}, Hamburg 1997 
\bibitem{limits} P. Janot,  Proceedings of the {\it International Europhysics
                Conference on High Energy Physics}, Jerusalem 1997
\bibitem{higgs} A. Denner, R. Guth, W. Hollik, J.H. K\"uhn,
                \zp C {\bf 51} (1991) 695; \\
                J. Rosiek, \plb B {\bf 252} (1990) 135; \\
                M. Boulware, D. Finnell, \prd D {\bf 44} (1991) 2054
\bibitem{susy1}  G. Altarelli, R. Barbieri, F. Caravaglios,
                \plb B {\bf 314} (1993) 357; \\
                C.S. Lee, B.Q. Hu, J.H. Yang, Z.Y. Fang,
                {\it J. Phys.\ } G {\bf 19} (1993) 13; \\
                Q. Hu, J.M. Yang, C.S. Li, {\it Comm.\ Theor.\ Phys.\ }
                {\bf 20} (1993) 213; \\
                J.D. Wells, C. Kolda, G.L. Kane, 
                                 \plb B {\bf 338} (1994) 219; \\
                G.L. Kane, R.G. Stuart, J.D. Wells,
                \plb B {\bf 354} (1995) 350; \\
                M. Drees et al., \prd D {\bf 54} (1996) 5598; \\
                J. Ellis, G. Fogli. E. Lisi, \plb B {\bf 389} (1996) 321
\bibitem{susydelr} P. Chankowski, A. Dabelstein, W. Hollik, W. M\"osle,
                   S. Pokorski, J. Rosiek, \np B {\bf 417} (1994) 101; \\
                   D. Garcia,  J. Sol\`a, {\it Mod.\ Phys.\
                   Lett.\ } A {\bf 9} (1994) 211
\bibitem{susy3} D. Garcia, R. Jim\'enez, J. Sol\`a,
                \plb B {\bf 347} (1995) 309; B {\bf 347} (1995) 321;
                D. Garcia, J. Sol\`a, \plb B {\bf 357} (1995) 349; \\
                A. Dabelstein, W. Hollik, W. M\"osle, in
                {\it Perspectives for  Electroweak
                Interactions in $\epm$ Collisions},
                Ringberg Castle 1995,
                Ed.\ B.A. Kniehl, World Scientific 1995 (p.\ 345); \\
                P. Chankowski, S. Pokorski, \np B {\bf 475} (1996) 3
\bibitem{deboer} W. de Boer, A. Dabelstein, W. Hollik, W. M\"osle,
                 U. Schwickerath, \zp C {\bf 75} (1997) 627
\bibitem{drosusy} A. Djouadi, P. Gambino, S. Heinemeyer, W. Hollik,
                  C. J\"unger, G. Weiglein, 
                  \prl {\bf 78} (1997) 3626; 
                          -- hep-ph/9710438
\bibitem{ellisetal} J. Ellis, G. Ridolfi and F. Zwirner, \plb
                     B {\bf 257}   (1991) 83 
\bibitem{hunter}  H. P. Nilles, \it Phys. Rep. \rm \bf 110  \rm (1984) 1; \\
      H. E. Haber, G. Kane, \it Phys. Rep. \rm \bf 117  \rm (1985) 75; \\
      J. F. Gunion, H. E. Haber, \it Nucl.\ Phys.\  \rm B \bf 272  
                                                    \rm (1986) 1; \\
      \it Nucl.~Phys.\ \rm B \bf 402 \rm (1993) 567; \\
      J. F. Gunion, H. E. Haber, G. Kane, S. Dawson: \it
      The Higgs Hunter's Guide, \rm  Addison-Wesley 1990.
\bibitem{hoang} R. Hempfling, A. Hoang, \plb B {\bf 331} (1994) 99; \\
                M Carena, M. Quiros, C. Wagner, \np B {\bf 461} (1996) 407
\bibitem{heinemeyer} S. Heinemeyer, {\it 2-Loop Calculations in the MSSM}, 
                  these Proceedings
\bibitem{jeger}    S. Eidelman, F. Jegerlehner,
                   \zp C {\bf 67} (1995) 585; \\
                    H. Burkhardt, B. Pietrzyk,
                    \plb B {\bf 356} (1995) 398
\bibitem{hollik} W. Hollik, {\it Review Status of the Standard Model},
                these Proceedings 
\bibitem{schwick} U. Schwickerath, Proceedings of the {\it XVI
                International Workshop on Weak Interactions and
                Neutrinos (WIN'97)}, Capri 1997
\bibitem{cleo} CLEO-Collaboration, R.~Ammar et al., 
                {\it Phys.~Rev.~Lett.}~{\bf 74} (1995) 2885; 
                M. Feindt, Proceedings of the {\it International Europhysics
                Conference on High Energy Physics}, Jerusalem, August 1997
                (to appear)
\bibitem{barbieri} 
              R. Barbieri, G. Giudice, {\it Phys.~Lett.\ }
               B {\bf 309} (1993) 86; \\
              R. Garisto, J.N. Ng, {\it Phys.~Lett.\ }
                B {\bf 315} (1993) 372\rm; \\
              S. Bertolini, F. Borzumati, A.Masiero,  G. Ridolfi,
              \it  Nucl.~Phys.\ \rm B 
               {\bf 353} (1991) 591 {\em and references therein}; \\
               N. Oshimo, \it  Nucl. Phys. \rm B {\bf 404} (1993) 20; \\
              S. Bertolini, F. Vissani,  \it Z. Phys. \rm C \bf 67 
                                       \rm (1995) 513, 1995

\bibitem{erler} D.M. Pierce, J. Erler, hep-ph/9708374 
 
 

\end{thebibliography}
